\documentclass[pre,amsmath,twocolumn]{revtex4}
\usepackage{graphicx} 
\usepackage{epsfig}

\begin{document}
\title{Renormalized Jellium model for charge-stabilized colloidal
suspensions
}
\author{Emmanuel Trizac$^1$\thanks{E-mail:
    \email{Emmanuel.Trizac@th.u-psud.fr}} and  Yan Levin$^{2,1}$\thanks{E-mail: \email{levin@if.ufrgs.br}}}

\affiliation{$^1$ Laboratoire de Physique Th\'eorique, UMR CNRS 8627, Universit\'e
de Paris XI, B\^atiment 210, F-91405, Orsay Cedex, France}

\affiliation{$^2$ Instituto de F\'{\i}sica, 
  Universidade Federal do Rio Grande do Sul \\ 
  Caixa Postal 15051, CEP 91501-970, Porto Alegre, RS, Brazil}


\date{\today}

\begin{abstract}

We introduce a renormalized Jellium model 
to calculate the equation of state for charged colloidal suspensions.
An almost perfect agreement with Monte Carlo simulations is found.
Our self-consistent approach naturally allows to define the effective charge of 
particles {\em at finite colloidal density}.  Although this quantity may differ
significantly from its counterpart obtained from the standard
Poisson-Boltzmann cell approach, the osmotic 
pressures for both models are in good agreement.  We argue that by
construction, the effective charge obtained using the Jellium
approximation is more appropriate to the study of colloidal interactions.
We also discuss a possibility of a fluid-fluid critical point and
show how the new equation of state can be used to
shed  light on the surprising results found in recent sedimentation
experiments.

\end{abstract}

\maketitle

In spite of the great effort invested in trying to 
understand the
phase stability of colloidal suspensions, our knowledge of
these complex systems is still quite rudimentary. It is curious
to compare this situation with an earlier debate concerning 
the nature, or even the possibility of the 
liquid-gas phase separation in 
symmetric electrolytes. Now this debate is almost over, and
the phase structure of a symmetric
1:1 electrolyte is well elucidated, although the 
universality class of the critical point is still being discussed. 
The Coulombic criticality 
---to distinguish it from the  solvophobic criticality---
cannot be observed in water and 
organic solvents of low dielectric permittivity must be used \cite{Fis94}. 
The phase separation  results from the strong 
electrostatic correlations between the
cations and anions of electrolyte \cite{Lev02}.

The phase stability of charged colloidal suspensions, on the other
hand, is far from being well understood 
\cite{vRoij,Linse,Warren,Diehl,Deserno,Trizac,Tamashiro,Dufreche}.
The strong charge and size asymmetry between the macroions and the
microions present in a suspension makes it very difficult
to apply to these systems the traditional tools of the liquid state
theory. Nevertheless, in view of results on 
symmetric 1:1 electrolytes, it would not be very 
surprising if  colloidal suspensions inside a solvent of sufficiently low
dielectric permittivity $\epsilon$ also presented a gas-liquid phase transition.  
Indeed such an instability has been observed in recent Monte Carlo
simulations, see for example \cite{Lobaskin,Linse} and references therein.  
What is much more surprising 
is that there are
some experimental indications of an instability even in 
aqueous suspensions containing {\it only} 
monovalent counterions. Theoretical 
estimates of the strength of electrostatic 
correlations, for aqueous suspensions suggest that they
should be too ``hot'' for 
an instability to set in. Nevertheless the experimental situation
remains unclear \cite{Bla}. 

A number of theories have been proposed to address this
unsettling experimental situation. A major drawback of most of these
approaches is that they rely on  uncontrolled approximations which
have not been fully tested. 
However, numerical solution of the
full non-linear Poisson-Boltzmann (PB) equation 
inside a spherical Wigner-Seitz (WS)
cell \cite{Marcus} finds no indication of 
any thermodynamic instability \cite{Deserno}.  
Of course, one can rightly question the reliability of the  
Wigner-Seitz cell PB model for the study of a fluid
phase of a highly disordered suspension.  On the one hand, since the
main contribution to the osmotic pressure inside an aqueous suspension
comes from the polyion-microions interactions, 
the cell might not be such a bad approximation.  On the other hand, 
the cell fails to properly account for the colloid-colloid correlation,
but these might not be of much importance for aqueous suspensions
with monovalent counterions.
Nevertheless, while the cell model is a good approximation for
dense colloids, one should be very careful in extrapolating 
its findings to highly dilute
suspensions.  Clearly there is an urgent need for an accurate
theory which will not rely on the cell approximation
and which
would be relevant for the study of colloidal phase stability.  
In this Letter we shall present such a theory.  
Our approach is similar to the Jellium approximation much used
in the solid state physics.
 
Consider an aqueous suspension of colloidal particles of 
charge $-Zq$ and radius $a$ in contact with a reservoir of monovalent
salt at concentration $c_s$ and electrostatic potential $\phi_r=0$
($q$ is the elementary charge).
The number of counterions and coions inside the suspension
is determined by the thermodynamic equilibrium.
While the colloidal particles are more or less  
uniformly distributed throughout the 
solution ---we are mainly interested in the small density regime---, 
the positions of counterions and coions are strongly
correlated with the positions of polyions.  As a leading
order approximation we can, therefore, take the polyion-polyion
correlation function to be $g_{pp}=1$ \cite{Beresford}
while the exact polyion-counterion and polyion-coion 
correlation functions are  
$g_{p\pm}=e^{-\beta w_{\pm}(r)}$, where $w_{\pm}(r)$  are 
the polyion-counterion and the polyion-coion 
potentials of mean force and $\beta=1/(k_B T)$ is the inverse temperature.

Choosing the coordinate system in such a
way that it is centered on top of one of the colloidal particles,
the electrostatic potential 
satisfies the Poisson equation 
\begin{equation}
\label{0}
\nabla^2 \phi=-\frac{4 \pi}{\epsilon} \rho_q(r) \;.
\end{equation}
The charge density is  
$\rho_q(r)=-\rho_{\text{back}}+q \rho_+(r)-q \rho_-(r)$, 
where, $\rho_{\text{back}}=Z_{\text{back}} q \rho_p$,
and $\rho_p$ is the mean density of colloids inside suspension.
The background charge is excluded from the colloidal interior.
Naively one can suppose that $Z_{\text{back}}=Z$. This, however, is not correct
and the bare charge must be renormalized in such a way as to lead to a 
self-consistent solution of Eq. (\ref{0}), as discussed below.

Approximating the potential of mean force by the  electrostatic
potential, the local concentration of counterions and coions
inside a suspension is,
\begin{equation}
\label{1}
\rho_\pm(r) \, = \, c_s \, e^{\mp\beta q \phi(r)} .
\end{equation}
It is important to keep in mind that in order for suspension
to be neutral, the electrostatic potential at infinity (bulk) cannot
vanish but must saturate to a
value $\phi(\infty)=\phi_D$ given by 
$c_s \sinh[\beta q \phi_D] = \rho_p Z_{\text{back}}$.
There exists, therefore,
an electrostatic potential difference 
between the suspension
and the salt reservoir.
In the biophysics literature, this potential difference is known
as the Donnan potential, 
and is partially responsible for
the biological cell trans-membrane potential.
 
Far away from the colloidal
surface, the electrostatic potential 
reduces to  the familiar Debye-H\"uckel 
expression
\begin{equation}
\label{3}
\phi(r)=\phi_D-\frac{Z_{\text{eff}}\, q}{\epsilon (1+\kappa a) \,r } 
\, e^{-\kappa(r-a)}\;,
\end{equation}
where $\kappa^2 = 4 \pi \lambda_B [\rho_+(\infty)+\rho_-(\infty)]$
and $\lambda_B = \beta q^2/\epsilon$ denotes the Bjerrum length.
The value of $Z_{\text{eff}}$ is determined self-consistently from 
the  numerical solution of Eq.~(\ref{0}) 
so that $Z_{\text{eff}}(Z,Z_{\text{back}},c_s)=Z_{\text{back}}$.
This  renormalization  of background charge is a consequence of
counterion condensation. 
It is important to keep in mind that for highly charged colloidal
particles, $Z_{\text{eff}}$ is {\it not} equal to the bare colloidal charge.
Furthermore, using the contact theorem and the vanishing of
the electric field 
as $r\to \infty$,
the osmotic pressure within the suspension takes
a simple form
\begin{equation}
\label{4}
\beta P=\rho_p + \rho_+(\infty)+\rho_-(\infty)-2 c_s=
\rho_p+\sqrt{Z_{\text{eff}}^2 \rho_p^2+4c_s^2}-2 c_s\;.
\end{equation}
It is a nice feature of the Jellium
approximation that once the effective charge is determined, the Debye
length $1/\kappa$  and 
the osmotic pressure both follow directly.  
\begin{figure}[htbp]
\includegraphics[width=8cm]{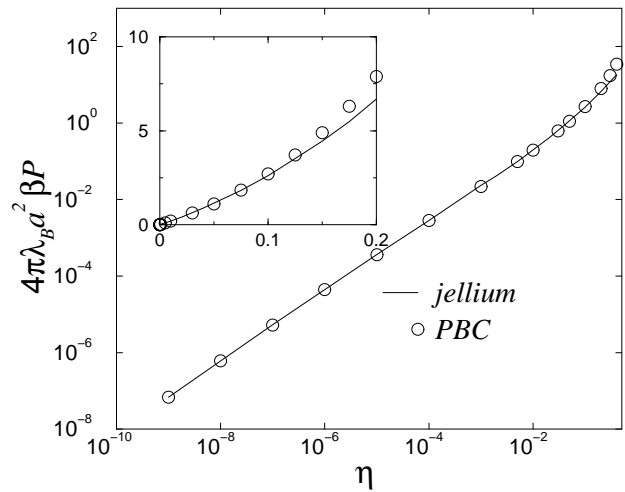}
\caption{Pressure-volume fraction isotherms 
within the Poisson-Boltzmann cell model (PBC)
and the renormalized Jellium approximation for $c_s=0$
(deionized situation). The packing fraction is defined
as $\eta=4 \pi \rho_p a^3/3$.
The inset shows the same data on the linear scale.}
\label{fig1}
\end{figure}
In Fig. \ref{fig1} we compare the
osmotic pressures calculated using the renormalized 
Jellium approximation
to the ones obtained within the WS cell
(for both approaches, the effective charges and pressures become
independent of $Z$ when the latter is large enough,
approximately $Z>20 a/\lambda_B$; Fig. \ref{fig1} has been
obtained under this condition of saturation, which is usually met
in colloidal suspensions).  
A surprisingly
good agreement is found between the two theories,
with a discrepancy only for volume fractions $\eta >0.15$.
In Fig \ref{fig2}, the 
osmotic pressure calculated in the
renormalized Jellium model is compared to the
results of recent Monte Carlo simulations \cite{Linse}
(where a model system of charged spherical macroions
and point counterions interacting solely through
hard sphere and Coulomb forces has been considered). The agreement 
is excellent, and justifies the neglect of colloid-colloid correlations
in our treatment. Clearly, for high packing fractions (namely $\eta>0.1$)
colloid-colloid correlations become important and should invalidate 
our approach.
\begin{figure}[htbp]
\includegraphics[width=8cm]{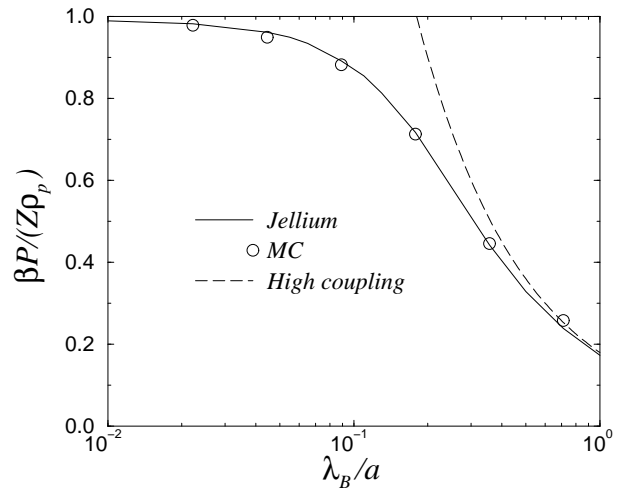}
\caption{ Comparison between the osmotic 
coefficient $\beta P/(Z \rho_p)=Z_{\text{eff}}/Z$  calculated using 
Monte Carlo (MC) simulations \cite{Linse} and the
renormalized Jellium model for  $c_s=0$, $Z=40$, $\eta=0.00125$.
At large couplings, we predict $\beta P/(Z\rho_p) \approx 0.1802\, a/\lambda_B $, 
shown as the dashed curve. }
\label{fig2}
\end{figure}

\begin{figure}[htbp]
\includegraphics[width=8cm]{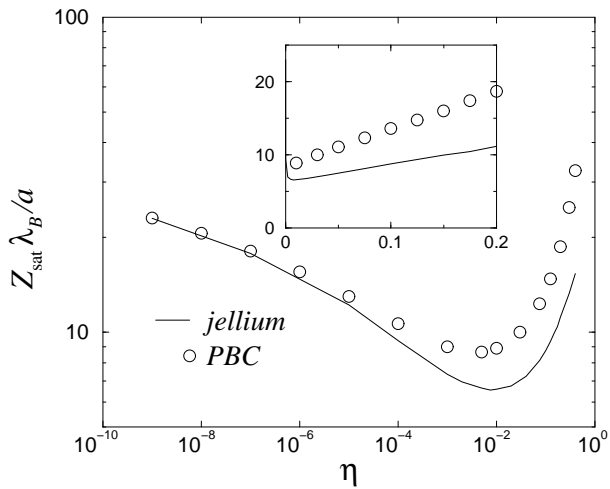}
\caption{Effective values of colloidal 
charges calculated with the  Poisson-Boltzmann cell model \cite{Alexander}
and the renormalized Jellium approximation for $c_s=0$. 
Inset shows the
same data but on a linear scale.}
\label{fig3}
\end{figure}

In Fig. \ref{fig3},
the  effective charges calculated using the cell model \cite{Alexander}
and the renormalized Jellium theory
are compared.  At very low volume fractions ($\eta<10^{-5}$)
the saturated effective charge of a salt-free suspension 
is in perfect agreement with the value of saturated effective
charge found within the WS cell and can be approximated
by a simple equation,
\begin{equation}
\label{4a}
Z_{\text{sat}} \approx 
\frac{a}{\lambda_B}\left[\delta-\gamma \ln (\eta)\right]\;,
\end{equation}
where $\gamma \approx 1$
and $\delta\approx 2$.
For volume fractions $\eta>10^{-4}$, there is a fairly strong
disagreement between the effective charges
predicted by the two theories, even if both models are in  
very good agreement for the value
of the osmotic pressure. 

If the effective charge is to be used 
to study the structural properties of colloidal suspensions
(such as the structure factors or any other quantity 
requiring the knowledge of the 
effective interaction potential between two polyions),
we argue that the Jellium effective charge is the more relevant quantity. 
The reason for this
is that within the Jellium approximation, 
two colloids at large separations $R$ interact
by the usual DLVO potential, as follows after some algebra
from integrating the stress tensor over a colloid's surface 
(see also \cite{Lev02})
\begin{equation}
\label{5}
V(R)=\frac{Z_{\text{eff}}^2 q^2}{ \epsilon (1+\kappa a)^2\, R } e^{\kappa (2a-R)}\;.
\end{equation}
This is not the case for two colloids inside a WS cell for which
the interaction potential is a very 
complicated function of separation and can only be calculated
numerically. 
In fact, for separations larger than the inter-particle distance, it
is even difficult to properly define what one means
by an interaction potential between colloids within the WS
formalism, since the overall charge neutrality 
results in a vanishing electrostatic interaction between the two cells.

To assess the quantitative validity of our approach, we compare
our (saturated) effective charges to those deduced in recent 
experiments using a torsional resonance spectroscopy \cite{Wette}. 
For colloidal volume fractions in the interval 
$10^{-3}<\eta<3.10^{-2}$,
$Z_{\text{sat}} \lambda_B/a$ was found to be close to $6$ \cite{Wette}.
This is in much better agreement with the 
renormalized Jellium model, which finds for these volume fractions
$Z_{\text{sat}} \lambda_B/a$ varying between
$6.7-7.0$, as compared to the WS cell prediction of $8.7-9$.  
We also note that a value close to 6 has been reported 
for this density regime in the Monte Carlo 
study presented in Ref \cite{Stevens}.

We also emphasize that within the Jellium model, the inverse screening length
$\kappa$ is naturally related to the effective salt density
[see the definition below Eq. (\ref{3})]. This should be contrasted with
the cell model, for which there is no simple connection between
the two quantities.

The Jellium model predicts stability of a charged 
colloidal suspension
against a fluid-fluid phase separation.  
This reinforces the cell picture, where no
instability is found. Our approach, however, neglects microionic correlations,
which become important at high electrostatic 
couplings \cite{Rouzina,Schklovskii,Lev02},
more precisely when $\Gamma=v^2\lambda_B/d$ exceeds
a threshold close to 2, $d$ being a characteristic distance between
microions in the electric double-layer and $v$ their valency.
In practice, however,  ionic hydration puts a lower bound to $d$, which
prevents the
high coupling regime from ever being reached
in water with monovalent counterions. 
Under these conditions our approach should, therefore, 
be quite reliable. Alternatively, 
when dealing with point particles, $d$ may be estimated as 
$d \simeq (4\pi a^2 v/Z)^{1/2}$ \cite{Rouzina,Lev02}.
For the Monte Carlo simulations reported in Fig. \ref{fig2}, 
$\Gamma=2$ therefore corresponds to
an instability threshold
$\lambda_B/a \simeq 1.1$.  Beyond this point, the microionic
correlations destabilize the system and lead to a fluid-fluid separation
\cite{Lobaskin,Linse}.
Nevertheless, the pressures obtained within the 
renormalized Jellium approximation are in very good agreement with the
Monte Carlo simulations, even at fairly large electrostatic couplings
(in Fig. \ref{fig2}, we have included the highest coupling
for which the pressure was computed in \cite{Linse}). 
We therefore expect that the Jellium approximation, suitably corrected
to include the counterion induced attraction between the colloidal
particles present at strong couplings, might be sufficient to account for
the thermodynamic instability in suspensions containing multivalent
ions.  Work along these lines is in progress.

Finally, it is interesting to speculate how the renormalized Jellium
model can help to shed new light on the problem of sedimenting colloidal 
dispersions \cite{Piazza,JPH,Simonin,Lowen,Tellez}.
Recently, Philipse and Koenderink (PK) \cite{PK} observed  strongly
inflated concentration profiles for charged mono-dispersed colloidal
particles in absolute ethanol.  Use of ethanol, instead of water,
allows to produce highly deionized suspensions with
``salt'' concentrations on the order of $10^{-9}$ M. The renormalized
Jellium predicts that for this salt concentration, an
infinitely dilute colloid would have an  effective
charge of $Z_{\text{eff}} \approx 20 \,a/\lambda_B$.   It
is difficult to know what is precisely the bare charge 
of colloidal particles
inside ethanol, however since the Bjerrum length
in ethanol is $\lambda_B \approx 2.3$ nm, a fairly small bare charge
on the order of a few hundred electrons is enough to
place the particles (with radius $a\simeq 90\,$nm \cite{PK})
in the saturation regime. 
Using this observation we can partially account 
for the observations of PK. 

Static equilibrium of a suspension in a gravitational field requires
that
\begin{equation}
\label{6}
\frac{d P}{d x}=- m g \rho_p\;,
\end{equation}
where $m$ is the colloidal mass (corrected for buoyancy), $g$ the 
gravitational acceleration, and $x$ the vertical displacement. 
It is convenient
to define the gravitational length as 
$l_g=1/(m g \beta)$, which for the experiments of PK is 
$l_g \approx 0.2\,$mm \cite{PK}. 
Substituting the expression for the
osmotic pressure Eq.(\ref{4}) into Eq.(\ref{6}) and using the dependence
of the saturated effective charge on the colloidal volume fraction,
Eq.(\ref{6}) can be integrated.
We find that for colloidal volume fractions $\eta>\eta_1 = 10^{-5} $
(i.e. in the counterion dominated regime): 
\begin{equation}
\label{7}
\ln(\eta/\eta_0)-\frac{1}{2} \left[\ln^2(\eta)-\ln^2(\eta_0)\right]=
\frac{\lambda_B x}{a l_g}\;,
\end{equation}
where $\eta_0$ is the reference volume fraction at the $x$ origin. 
This can be
taken as $\eta_0 \approx 0.01$, the point at which
the hard-core effects are completely negligible, 
and one can be sure to
be looking at the dilute tail of the sedimenting profile. Note that
for small volume fractions, $\eta \sim \exp(-\sqrt{2\lambda_B x/a l_g})$
unlike the simple exponential barometric law.
For $\eta <\eta_\times\approx 5 \times 10^{-8}$, 
the salt resulting from solvent dissociation
dominates over counterions, and we recover
the usual barometric law $\eta(x) \sim \exp(-x/l_g).$
Results concerning the crossover regime $\eta_\times<\eta<\eta_1$
will be published elsewhere.
It is interesting to estimate the extent of the sedimentation
profile predicted by the renormalized Jellium model.  Since the
concentrations of colloids on the order of $\eta_m\approx10^{-5}$ are still
detectable, we find that the profile extends distance
$x \sim a l_g \ln^2(\eta_m)/(2\lambda_B) \approx 2600\, l_g$.
For the colloids used by PK, this is almost $50$ cm! 
Hence, the observed inflation of the sedimentation profile.

To conclude, we have proposed a self-consistent renormalized  
non-linear Jellium model
that constitutes an alternative to the widely used PB cell approach
for charged colloidal suspensions.
Surprisingly, the equations of state within the two theories turn out to be
very close, even though that the approximations 
involved to account for finite colloidal density are 
very different (finite cell against
a renormalized background). Our results
point to the relevance of the cell picture
even at extremely low densities.
This reinforces the argument that a fluid-fluid phase instability
is impossible, for highly charged colloids in water at room temperature,
as long as the microions are monovalent [which allows to neglect
microionic correlations, thereby identifying the potential of mean force
with the electrostatic potential, see Eq. (\ref{1})]. 
Our approach also allows to define in a natural way, not only 
the effective charge of the macroions, but also the effective screening
length; these quantities are experimentally and conceptually more relevant
than those obtained within the cell approach. Finally, 
our equation of state for deionized systems helps in understanding
recent sedimentation experiments where ``anomalous'' density profiles have
been reported.

This work was supported in part by the Brazilian agencies
CNPq and FAPERGS and by the french CNRS. 


\end{document}